\documentclass[preprint,onecolumn]{revtex4-1}

\usepackage{graphicx}
\usepackage{amssymb}
\usepackage{amstext}
\usepackage{amsbsy}
\usepackage{amsmath}

\begin{document}

\title{Multiple steadily translating bubbles in a Hele-Shaw channel}

\author{Christopher C. Green$^{1}$ and Giovani L. Vasconcelos$^{1,2}$ }\address{$^{1}$Department of Mathematics, Imperial College London, 180 Queen's Gate, London SW7 2AZ, United Kingdom\\
$^{2}$Departamento de F\'{\i}sica,  Universidade Federal de Pernambuco, 50670-901, Recife, Brazil}




\begin{abstract}
Analytical solutions  are constructed for an assembly of any finite number of bubbles in steady 
motion in a Hele-Shaw channel. The solutions are given in the form of 
a conformal mapping from a bounded multiply connected circular domain to the flow region exterior to the bubbles. The mapping is written as the sum of two analytic functions---corresponding to the complex potentials in the laboratory and co-moving frames---that map the circular domain onto respective degenerate polygonal  domains. These functions are   obtained using the generalised Schwarz-Christoffel formula for multiply connected domains in terms of the Schottky-Klein prime function. Our solutions are very general in that no symmetry assumption concerning the geometrical disposition of the bubbles is made. Several examples for various bubble configurations are discussed.
\end{abstract}

\maketitle

\section{Introduction}

Many free boundary problems naturally arise from the consideration of different types of Hele-Shaw systems. A Hele-Shaw system is one where two viscous fluids (typically with one fluid much less viscous than the other)
are sandwiched between two closely spaced parallel plates so as to produce a flow that is essentially two-dimensional.  Hele-Shaw flows in various geometries have been extensively studied over the years, and 
many processes in physics involving the evolution of interfacial boundaries, such as dendritic crystal growth, direct solidification, and fluid displacement in porous media, can be modelled mathematically (under certain assumptions) as a free boundary problem of the Hele-Shaw type. This diverse array of free boundary problems 
has a plethora of analytical solutions and a wide range of mathematical  methods can be used to solve them (Pelc\'e \cite{Pelce}, Gustafsson \& Vasil'ev   \cite{Gustafsson}). 
The models defining these free boundary problems also go by the name of Laplacian growth processes (Mineev-Weinstein, Putinar \& Teodorescu \cite{Mark1}) because the governing field equation in the viscous fluid region is Laplace equation and the evolution of the fluid interfaces is governed through surface derivatives of this field. In the case of Hele-Shaw bubbles, the flow is governed by Darcy's law and the bubble interfaces evolve with a velocity 
proportional  to the local gradient of the fluid pressure.

In this paper, we present a  general analytical solution for the problem of multiple bubbles  steadily translating along a Hele-Shaw channel when surface tension effects  on the bubble boundaries are neglected.
The mathematical problem to be solved belongs to a class of free boundary problem defined in a multiply connected domain which is related to a special class of scalar Riemann-Hilbert problem (Gakhov \cite{Gakhov}) recently considered by Crowdy \cite{DarrenRHpaper}.
Here, however, we shall use a more direct approach which allows us to reduce our original free boundary problem to  two ``fixed boundary'' problems which can  be more easily solved. 
More specifically, we introduce a conformal mapping from a circular domain in an auxiliary complex $\zeta$-plane to the physical flow domain (i.e., the exterior of the bubbles bounded by the channel walls) in the complex $z$-plane and   show that this  
mapping can be written as the sum of two  analytic functions,  corresponding  to the complex potentials in the laboratory frame and in the co-moving frame, respectively.  Each one of these functions maps the circular domain onto a multiply connected degenerate polygonal domain and hence can be constructed using a  variation of the generalised Schwarz-Christoffel mapping for multiply connected domains 
recently obtained by Crowdy \cite{BoundedMCSC,UnboundedMCSC} in terms of the Schottky-Klein prime function. 
Our final expression for the conformal map revealing the bubble  shapes will be given as an explicit indefinite integral whose integrand consists of  products of Schottky-Klein prime functions and their derivatives.

There are several prior results pertaining to steady multiple bubbles in Hele-Shaw systems which we wish to survey to motivate the free boundary problem  considered in this paper. An important assumption that is made in each of these works is the exclusion of surface tension effects. From a theoretical standpoint, this makes the problem analytically tractable and allows for exact solutions to be found. Taylor \& Saffman \cite{TS} found an exact solution for a single bubble in a channel with reflectional symmetry about the channel centreline. Tanveer \cite{TanveerBub} was able to generalise this solution using elliptic function theory to describe a single asymmetric bubble in the channel. Vasconcelos \cite{Giovani1} reported exact solutions for a finite number of steadily translating bubbles in a Hele-Shaw channel. He considered two symmetrical classes of solutions for bubbles which are either symmetrical about the channel centreline or which possess fore-and-aft symmetry, and derived Schwarz-Christoffel type formulae for the conformal mappings determining the bubble interfaces. Adopting a similar approach, Silva \& Vasconcelos \cite{GiovaniSilva}   have found exact solutions for a doubly periodic array of multiple symmetrical bubbles, with Schwarz-Christoffel methods again proving to be fruitful. More recently, these authors (Silva \& Vasconcelos \cite{GiovaniSilva2}) obtained a solution for a stream of asymmetric bubbles in a Hele-Shaw by deploying the Schwarz-Christoffel formula for doubly connected domains. Vasconcelos \cite{Giovani2,Giovani3} has also found families of exact solutions for various infinite streams of bubbles in the Hele-Shaw system.

Most relevant to our present free boundary problem is the work of Crowdy \cite{CrowdyUnboundedHSbubs}  who found analytical solutions determining the shapes of any finite number of steadily translating bubbles
in an unbounded Hele-Shaw cell. He derived analytical expressions for both the complex potential and the conformal map from a bounded multiply connected circular domain to the exterior of the bubble assembly by using the Schottky-Klein prime function.
The solutions presented herein  can be viewed as the generalisation to the channel geometry of the solutions found by Crowdy \cite{CrowdyUnboundedHSbubs}  for multiple bubbles in an unbounded  Hele-Shaw cell. Our solutions thus account for the effect of the two channel walls which greatly influence the nature of the free boundary problem. An earlier attempt at solving this problem was made by Crowdy \cite{wrongPaper},
but it was subsequently found that part of the argument used there was not valid  (Crowdy, private communication). Herein the problem is solved by employing the generalised Schwarz-Christoffel formula. 

\section{Problem formulation}
\label{sec:2}

\subsection{The complex potentials}
\label{sec:2a}

\begin{figure}[t]
\centering\includegraphics[width=1.0\textwidth]{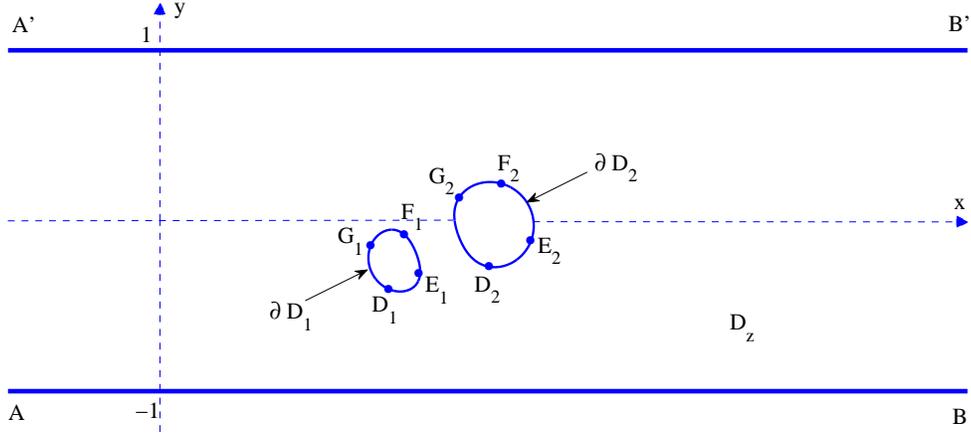}
\vspace{-0.5cm}
\caption{Schematic of a Hele-Shaw channel of width 2 containing an assembly of bubbles steadily translating with constant velocity $U$ parallel to the $x$-axis. The far field fluid velocity is assumed to be $V=1$, and the shapes of the bubble boundaries are to be determined.}
\label{BubblePhysDom}
\end{figure}

We consider the problem of $M$ finite-area bubbles translating  uniformly with speed $U$ parallel to the $x$ axis in a Hele-Shaw channel filled with an incompressible viscous fluid.  Without loss of generality, we will assume that the Hele-Shaw channel has a width equal to 2 and that  the viscous fluid (outside the bubbles) has a uniform speed $V=1$ in the far field; see Fig.~\ref{BubblePhysDom} for a schematic. Our model will be centred around some simplifying assumptions in order to render the problem analytically tractable. We shall assume that the fluid inside the bubbles (say, air) has negligible viscosity so that the pressure inside each bubble is constant. We also neglect surface tension effects, which implies that the viscous fluid pressure will have a constant value on each bubble boundary. We assume furthermore that the Hele-Shaw channel is horizontally placed so that the effects of gravity can be neglected.  Finally, we shall neglect three-dimensional thin film effects. 

Under the assumptions above, the motion of the viscous fluid in our Hele-Shaw channel is governed by Darcy's law:
\begin{equation}
\mathbf{v}=\nabla \phi
\label{bfvqn}
\end{equation}
where  the velocity potential $\phi(x,y)$ is given by
\begin{equation}
\phi=-\frac{b^{2}}{12\mu} p.
\end{equation}
Here $\mathbf{v}$ is the averaged fluid velocity across the channel, $p$ is the viscous fluid pressure, $b$ is the gap between the plates, and $\mu$ is the fluid viscosity. From  the incompressibility condition, $\nabla \cdot \mathbf{v}=0$, it follows that   $\phi$ satisfies Laplace equation, $\nabla^{2}\phi=0$. It is therefore natural to formulate the free boundary problem to be solved 
in the complex $z$-plane, where $z=x+\mathrm{i}y$.

\begin{figure}[t]
\centering\includegraphics[scale=0.4]{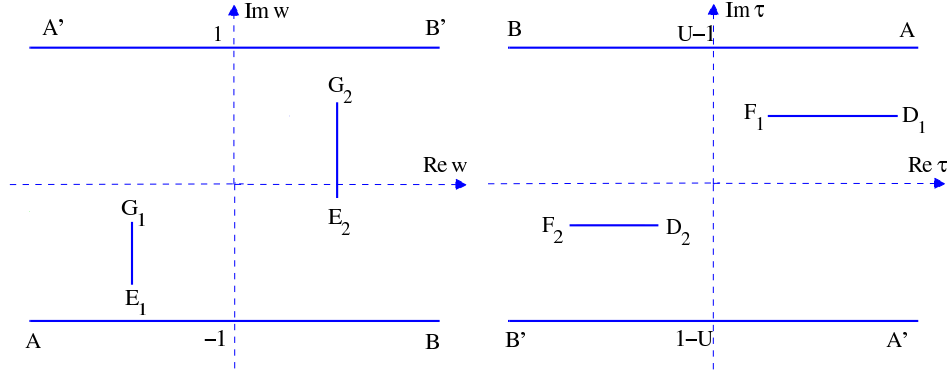}
\caption{Schematic showing the flow domains in the $w$-plane (left panel) and in the $\tau$-plane (right panel), representing  the complex potentials in the laboratory and co-moving frames, respectively. The slits in both cases correspond to the bubbles in Fig.~\ref{BubblePhysDom}.}
\label{schems}
\end{figure}

Let us thus introduce  the complex potential, $w(z)$, 
defined by
\begin{equation}
w(z)=\phi(x,y) +\mathrm{i}\psi(x,y)
\label{Ffunclabo}
\end{equation}
where  $\psi$ is the stream function associated with the velocity potential $\phi$. 
Since the viscous fluid is assumed to have unit speed in the far-field, it follows that 
\begin{equation}
w(z) =z\qquad \mbox{for}\qquad |x|\to\infty.
\label{5pt81mU}
\end{equation}
We shall label the viscous fluid region by $D_z$ and the boundary of the $j$-th bubble by $\partial D_j$, $j=1,...,M$; see Fig.~\ref{BubblePhysDom}.  Apart from a simple pole at infinity, 
the complex potential  $w(z)$ is analytic everywhere in the viscous fluid region $D_{z}$ and must satisfy the following boundary conditions:
\begin{equation}
\mathrm{Im}[w(z)]=\pm 1 \qquad \mbox{for}\qquad y=\pm 1,
\label{bubStrem2}
\end{equation}
\begin{equation}
\mathrm{Re}[w(z)]=\textrm{constant}\qquad \mbox{for}\qquad z\in\partial D_{j}.
\label{wuxk}
\end{equation}
Condition (\ref{bubStrem2}) simply states that the channel walls, $y=\pm 1$, are streamlines of the flow, whereas (\ref{wuxk})  follows from the fact that pressure $p$ is constant on each bubble boundary.
From conditions (\ref{bubStrem2}) and (\ref{wuxk}) one sees that the flow domain in the complex potential $w$-plane consists of a horizontal strip of width 2 with $M$ vertical slits in its interior, where each slit  corresponds to a bubble in the $z$-plane; see left panel of Fig.~\ref{schems}.

Let us also introduce the function $\tau(z)$ corresponding to the complex potential in a  frame of reference co-travelling with the bubbles: 
\begin{equation}
\tau(z)=w(z)-Uz.
\label{refwtU}
\end{equation}
From (\ref{5pt81mU}) and (\ref{refwtU}), one immediately sees that the far-field behavior of $\tau(z)$ is 
\begin{equation}
\tau(z) =(1-U)z\qquad \mbox{for}\qquad |x|\to\infty,
\label{eq:tz1}
\end{equation}
which implies the following boundary conditions for $\tau(z)$ on  the channel walls:
\begin{equation}
\mathrm{Im}[\tau(z)]=\pm(1-U)\qquad \mbox{for}\qquad y=\pm 1.
\label{bubStrem3}
\end{equation}
Furthermore, in the co-travelling frame the bubble boundaries are necessarily streamlines of the flow,  thus
\begin{equation}
\mathrm{Im}[\tau (z)]=\textrm{constant}\qquad \mbox{for}\qquad z\in\partial D_{j}.
\label{bubStrem4}
\end{equation}
It then follows from (\ref{bubStrem3}) and (\ref{bubStrem4}) that the flow domain in the $\tau$-plane is a horizontal strip of width $2(U-1)$ with horizontal slits in its interior corresponding to the bubbles;  see right panel of Fig.~\ref{schems}.

\subsection{The conformal mapping}
\label{sec:2b}

\begin{figure}
\begin{center}
\hspace{-1.8cm}\includegraphics[scale=0.4]{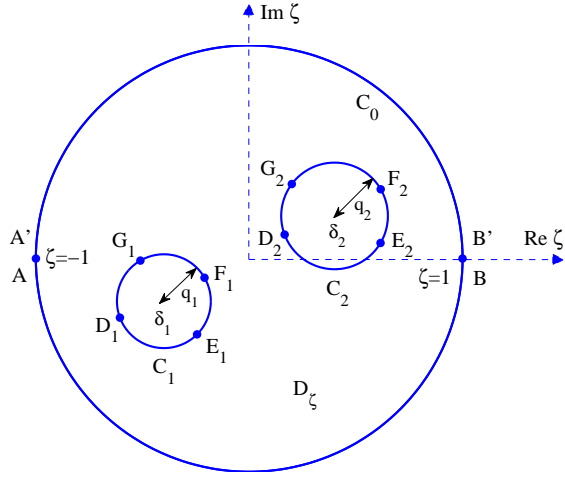}
\caption{The  circular domain $D_\zeta$ in the auxiliary complex $\zeta$-plane.
The case of a triply connected domain is illustrated, to be conformally equivalent to the physical domain in Fig.~\ref{BubblePhysDom}. The unit circle $|\zeta|=1$ maps to the channel walls, with $\zeta=\pm1$ being the preimages of $x=\pm\infty$  (the ends of the channel), respectively, and the circles $C_j$, $j=1,2$, map to the boundaries of the bubbles $\partial D_j$.}
\label{BubblePreimage}
\end{center}
\end{figure}

In order to determine the shapes of the bubbles, we shall construct the conformal map $z(\zeta)$ from a bounded $M+1$ connected circular domain $D_\zeta$  to the $M+1$ connected fluid region $D_z$ exterior to $M$ bubbles in the $z$-plane. We choose $D_\zeta$ to be  the unit circle $\zeta$-disc with $M$ smaller non-overlapping disks excised from it.
Label the unit circle by $C_0$ and label the $M$ inner circular boundaries as $C_1,...,C_M$, and let the centre and radius of $C_j$ be $\delta_j$ and $q_j$ respectively. 
A schematic of $D_\zeta$ is shown in Fig.~\ref{BubblePreimage} in the case where $M=2$ (triply connected). 
Let the unit circle $C_0$ map to the channel walls. 
This implies that the mapping function $z(\zeta)$ will necessarily have two logarithmic singularities on $C_0$. By the degrees of freedom afforded by the Riemann-Koebe mapping theorem (Goluzin \cite{Goluzin}), we can place these logarithmic singularities at $\zeta=\pm1$.  Let the interior circles $C_1,...,C_M$ map to the bubble boundaries $\partial D_1,...,\partial D_M$, respectively

Let us  define the functions $W(\zeta)$ and $T(\zeta)$  through the following compositions:
\begin{align*}
W(\zeta)=w(z(\zeta)) ,\\
 T(\zeta)=\tau(z(\zeta)).
\end{align*}
These functions 
must  be analytic in the circular domain $D_\zeta$ and satisfy appropriate boundary conditions on the circles $C_j$, as discussed next. 
From conditions (\ref{bubStrem2}) and (\ref{wuxk}) it follows that $W(\zeta)$ must be such that:
\begin{equation}
\mathrm{Im}[W(\zeta)]=\pm1, \quad \zeta \in C_0,
\label{RHprob1}
\end{equation}
\begin{equation}
\mathrm{Re}[W(\zeta)]=\textrm{constant}, \quad \zeta \in C_{j}, \quad j=1,...,M.
\label{RHprob2}
\end{equation}
Similarly, for $T(\zeta)$ one has from (\ref{bubStrem3}) and (\ref{bubStrem4}) that:
\begin{equation}
\mathrm{Im} [T(\zeta)]=\mp 1, \quad \zeta \in C_0,
\label{Bubblah1}
\end{equation} 
\begin{equation}
\mathrm{Im} [T(\zeta)]=\textrm{constant}, \quad \zeta \in C_{j}, \quad j=1,...,M.
\label{Bubblah2}
\end{equation}
In (\ref{RHprob1}) and (\ref{Bubblah1}) the upper and lower signs  correspond to the  upper and lower semicircles on the unit circle,
respectively. 

Now,  from (\ref{refwtU}) it follows that the mapping function $z(\zeta)$ can be written as
\begin{equation}
z(\zeta)=\frac{1}{U}\left[W(\zeta)-T(\zeta)\right].
\label{eq:z1}
\end{equation}
We have thus reduced our original free boundary problem to the more manageable task of computing two  analytic functions, 
$W(\zeta)$ and $T(\zeta)$, that  map the  circular domain $D_{\zeta}$  to  multiply connected slit domains which can be viewed as degenerate polygonal domains. 
These functions can then be readily obtained using a generalised Schwarz-Christoffel mapping for multiply connected polygonal domains, which is briefly reviewed in the next section.

\section{Generalised Schwarz-Christoffel mapping}
\label{sec:3}

Consider the function $z(\zeta)$ defined on the circular domain $D_\zeta$ by the following expression:
\begin{align}
z_{\zeta}(\zeta)
&=\mathcal{B} 
\frac{ \left[ \omega_\zeta(\zeta,1)\omega(\zeta, -1)-\omega_\zeta(\zeta,-1)\omega(\zeta,1)\right]}{\prod_{j=1}^{M}\omega(\zeta,\gamma_{1}^{(j)}) \omega(\zeta,\gamma_{2}^{(j)})}
\prod_{j=0}^{M} \prod_{k=1}^{n_j} \left[\omega(\zeta,a_{k}^{(j)})\right]^{\beta_k^{(j)}},
\label{eq:SC6}
\end{align}
where $\mathcal{B}$ is a complex constant and $\omega(\zeta,\gamma)$ is the Schottky-Klein  prime function associated with the domain $D_\zeta$. For a definition of the Schottky-Klein prime function and a discussion of some of its properties, see, e.g., Crowdy \cite{BoundedMCSC}.
The Schottky-Klein  prime function has deep connections with Riemann surface theory (Fay \cite{Fay}), but for the purposes of this paper it suffices to think of it as a special computable function (Crowdy \& Marshall \cite{cmft_compSK}). 
 
 It is shown elsewhere (Vasconcelos  \&  Green \cite{SC2013}) that the function defined in (\ref{eq:SC6}) conformally maps the circular domain $D_\zeta$ onto a bounded $(M+1)$-connected polygonal domain in the $z$-plane, where the unit circle $C_0$ is mapped to the outer boundary polygon $P_0$ and the interior circles $C_1,...,C_M$ are mapped to the inner polygonal  boundaries $P_j$, $j=1,...,M$, respectively. The points  $a_{k}^{(j)}\in C_j$, $k=0,1,...,n_j$, are the preimages of the vertices $z_k^{(j)}$ on polygon $P_j$, with $\pi \beta_k^{(j)}$ being the corresponding turning angles so that the internal angle at the respective vertex is $\pi\alpha_k^{(j)}=\pi (\beta_k^{(j)}+1)$, with $\alpha_{k}^{(j)}\in[0,2]$. The outer polygon can have one or more vertices at infinity, i.e.,  $|z_{k}^{(0)}|=\infty$, for some $k$, in which case  $\alpha_{k}^{(0)}\in[-2,0]$ because  the angle is measured with respect to the variable $\frac{1}{z}$ for $z\to\infty$. Inner polygons are also allowed to degenerate to line segments (i.e., slits).

The set of points $\{\gamma_1^{(j)}, \gamma_2^{(j)}\in C_j|j=1,...,M\}$ appearing in formula (\ref{eq:SC6}) are obtained by computing the zeros (on each inner circle) of the following equation:
\begin{equation}
\omega_\zeta(\zeta,1)\omega(\zeta, -1)-\omega_\zeta(\zeta,-1)\omega(\zeta,1)=0.
\label{eq:gj}
\end{equation}
This guarantees that the derivative $z_\zeta(\zeta)$ has no poles at the points  $\zeta=\gamma_1^{(j)}$ and $\zeta=\gamma_2^{(j)}$, so that the only singularities of $z(\zeta)$ in $D_\zeta$ are  the appropriate branch points at $\zeta=a_{k}^{(j)}$. Formula (\ref{eq:SC6}) is an alternative version of the generalised Schwarz-Christoffel mapping 
derived by Crowdy \cite{BoundedMCSC} that is more convenient to treat  multiply connected strip domains.

For later use, we  quote here an important property of the Schottky-Klein prime function (Crowdy \cite{BoundedMCSC}): the functions defined by 
\begin{equation}
F_j(\zeta;\zeta_1,\zeta_2)=\frac{\omega(\zeta,\zeta_1)}{\omega(\zeta,\zeta_2)}, \qquad \zeta_1,\zeta_2\in C_j, \quad j=0,1,...,M,
\label{eq:Fj}
\end{equation}
 have constant argument on all circles $C_k$. More specifically,  
 one has (see \cite{SC2013})
\begin{align}
\arg \left[F_j(\zeta;\zeta_1,\zeta_2)\right]= Q_{jk}(\zeta_1,\zeta_2) \qquad \mbox{for}\qquad \zeta\in C_k, \quad k=0,1,...,M,
\label{eq:argF}
\end{align}
where
 \begin{align}
Q_{jk}(\zeta_1,\zeta_2)= \pi \left[  v_k(\zeta_1)- v_k(\zeta_2)-v_j(\zeta_1)+ v_j(\zeta_2)\right]+\frac{1}{2}\left(\varphi^{(j)}_1-\varphi^{(j)}_2\right).
\label{eq:Q}
\end{align}
Here,  the functions $v_j(\zeta)$, $j=1,...,M$,  are  the $M$ integrals of the first kind associated with the domain $D_\zeta$. An algorithm for computing  $v_j(\zeta)$  can be found in \cite{cmft_compSK}. For convenience, we  have  defined $v_0(\gamma)\equiv 0$ and introduced the notation $\varphi^{(j)}_1=\arg\left(\zeta_1-\delta_j\right)$, with similar definition for $\varphi^{(j)}_2$. In the next section, we shall use the generalised Schwarz-Christoffel formula (\ref{eq:SC6}) and property (\ref{eq:argF}) to give an explicit construction of the complex potentials $W(\zeta)$ and $T(\zeta)$ defined above.

\section{The general solution}

\subsection{The function $T(\zeta)$}
\label{sec:4a}

Recall that the  mapping $\tau=T(\zeta)$ conformally maps the circular domain  $D_\zeta$  onto a strip domain in the $\tau$-plane where the unit circle $C_0$ is mapped to the strip boundaries, ${\rm Im}[\tau]=\pm (U-1)$, and the inner circles $C_j$ are mapped to horizontal slits; see Fig.~\ref{schems} (right panel).
This domain can alternatively be viewed as a multiply connected 
degenerate polygonal domain, where the outer polygonal boundary has only two edges that meet at ${\rm Re}[\tau]=\pm\infty$ and the inner polygons  degenerate to slits. The corresponding turning angle parameters are therefore given by  $\beta_1^{(0)}=\beta_2^{(0)}=-1$ at the two vertices at infinity and   $\beta_1^{(j)}=\beta_2^{(j)}=1$, $j=1,...,M$, at the end points of the slits. Following previous notation, we   label by $a_1^{(0)}, a_2^{(0)}\in C_0$  the preimages in the $\zeta$-plane of the end points of the $\tau$-strip  and  by  $a_{1}^{(j)}, a_{2}^{(j)}\in C_j$, $j=1,...,M$, the preimages of the slit end points.
Now recall  that the preimages in the $\zeta$-plane of the channel left and right ends (which correspond to  ${\rm Re}[\tau]=\pm\infty$) have been chosen to be $\zeta=\mp1$, and so we have $a_1^{(0)}=-1$ and $a_2^{(0)}=1$. 
Applying the Schwarz-Christoffel formula (\ref{eq:SC6}) to this case then yields
\begin{equation}
T_{\zeta}(\zeta)=\mathcal{B} \left[\frac{\omega_{\zeta}(\zeta,1)\omega(\zeta,-1)-\omega_{\zeta}(\zeta,-1)\omega(\zeta,1)}{\omega(\zeta,1)\omega(\zeta,-1)}\right]
\prod_{j=1}^{M} \frac{\omega(\zeta,a_{1}^{(j)})\omega(\zeta,a_{2}^{(j)})}{\omega(\zeta,\gamma_{1}^{(j)})\omega(\zeta,\gamma_{2}^{(j)})}.
\label{eq:Tz}
\end{equation}

As discussed in Sec.~\ref{sec:3}, the points $\{\gamma_{1}^{(j)}, \gamma_{2}^{(j)}\in C_j|j=1,...,M\}$, are obtained by  computing the solutions  of (\ref{eq:gj}). 
In this case,  because the slits are all horizontal,  
one can select  the points  $\zeta=\gamma_{1}^{(j)}$ and $\zeta=\gamma_{2}^{(j)}$   to be the preimages of the slit end points (see \cite{SC2013}), i.e.,
\begin{align}
\gamma_{k}^{(j)}=a_{k}^{(j)}, \qquad j=1,...,M,
\label{eq:ak}
\end{align}
so that the terms in the numerator and in the denominator  of the product appearing in (\ref{eq:Tz}) will precisely  cancel out. Hence (\ref{eq:Tz}) simplifies to 
\begin{equation}
T_{\zeta}(\zeta)=\mathcal{B} \left[\frac{\omega_{\zeta}(\zeta,1)\omega(\zeta,-1)-\omega_{\zeta}(\zeta,-1)\omega(\zeta,1)}{\omega(\zeta,1)\omega(\zeta,-1)}\right],
\label{Wzeta}
\end{equation}
which upon integration becomes
\begin{equation}
T(\zeta)=\frac{2(U-1)}{\pi}\ln \left[\frac{\omega(\zeta,1)}{\omega(\zeta, -1)}\right],
\label{compPot}
\end{equation}
where we have  set   $\mathcal{B} = 2(U-1)/\pi$ to account for the fact that width of
the strip in the $\tau$-plane is 
$2(U-1)$; see right panel of Fig.~\ref{schems}. It is worth noting that the mapping (\ref{compPot}) can  be  obtained directly from the properties of the Schottky-Klein prime function, see, e.g., Crowdy \cite{wrongPaper}. It is instructive however to demonstrate that it can  be recovered from the Schwarz-Christoffel mapping (\ref{eq:SC6}), as shown above.

\subsection{The function $W(\zeta)$}
\label{sec:4b}

As discussed in Sec.~\ref{sec:2a}, the complex potential $W(\zeta)$ maps the circular domain $D_\zeta$ onto a  multiply connected domain in the $w$-plane consisting of a horizontal strip of width 2 with $M$ vertical slits in its interior, as shown in Fig.~\ref{schems} (left panel). This slit strip domain is therefore of the same type as the flow domain in the $\tau$-plane discussed above, the only difference being that the slits are now vertical. 
Let us then denote by $\{b_1^{(j)}, b_2^{(j)}\in C_j| j=1,...,M\}$ the set of  preimages in the $\zeta$-plane of the end points of the slits in the $w$-plane. It should  be clear from the preceding discussion that the function $W_\zeta$ is given by
\begin{equation}
W_{\zeta}(\zeta)=\mathcal{C} \left[\frac{\omega_{\zeta}(\zeta,1)\omega(\zeta,-1)-\omega_{\zeta}(\zeta,-1)\omega(\zeta,1)}{\omega(\zeta,1)\omega(\zeta,-1)}\right]
\prod_{j=1}^{M} \frac{\omega(\zeta,b_{1}^{(j)})\omega(\zeta,b_{2}^{(j)})}{\omega(\zeta,\gamma_{1}^{(j)})\omega(\zeta,\gamma_{2}^{(j)})},
\label{Tzeta}
\end{equation}
where $\cal C$ is a complex constant. The modulus of $\cal C$ has to be chosen such that the width of the strip in the $w$-plane equals 2; see below. On the other hand,  the argument of $\cal C$ and the parameters $\{b_{1}^{(j)}, b_{2}^{(j)}|j=1,...,M\}$ are determined from the requirement that $C_0$ is mapped by $W(\zeta)$ to horizontal straight lines and that the inner circles map to vertical slits, as  shown next.

First, it is  convenient to rewrite (\ref{Tzeta}) in terms of the functions $F_j(\zeta;\zeta_1, \zeta_2)$ defined in (\ref{eq:Fj}):
\begin{equation}
W_{\zeta}(\zeta)=\mathcal{C} \left[\frac{d}{d\zeta}\log F_0(\zeta;1, -1)\right]\prod_{j=1}^{M} \prod_{l=1}^2 F_j(\zeta;b_{l}^{(j)},\gamma_{l}^{(j)}).
\label{Tzeta2}
\end{equation}
The requirement that $C_0$ is mapped by $W(\zeta)$ to horizontal walls means that
\begin{equation}
\mathrm{Im} \left[\frac{ dW}{d\theta} \right]=0 \quad \mbox{for}\quad \zeta=e^{\mathrm{i}\theta} ,
\label{DT1a}
\end{equation}
which implies that $\arg[dW]=n\pi$, for some integer $n$, as one traverses an infinitesimal  angle $d\theta$ on the unit circle. This in turn implies from (\ref{Tzeta2}) and (\ref{eq:argF}) that
\begin{equation}
\arg[\mathcal{C}]+\sum_{j=1}^{M} \sum_{l=1}^2 
Q_{j0}(b_{l}^{(j)},\gamma_{l}^{(j)})=n\pi ,
\label{DT1}
\end{equation}
where  $Q_{jk}(\zeta_1,\zeta_2)$ 
is given in (\ref{eq:Q}).
Similarly, the condition that the inner circles $C_k$ are mapped to vertical slits can be written as  
\begin{equation}
\mathrm{Re} \left[\frac{ dW}{d\theta} \right]=0 \quad \mbox{for}\quad \zeta=\delta_k+q_ke^{\mathrm{i}\theta}, \quad k=1,...,M,
\label{DT2a}
\end{equation}
which implies that $\arg[dW]=(m+1/2)\pi$, for some integer $m$,  as one traverses an infinitesimal  angle $d\theta$ on the inner circle $C_k$. From this condition and (\ref{Tzeta2}), it  follows that
\begin{equation}
\arg[\mathcal{C}]+\sum_{j=1}^{M} \sum_{l=1}^2 
Q_{jk}(b_{l}^{(j)},\gamma_{l}^{(j)})=\left(m+\frac{1}{2}\right)\pi, \qquad k=1,...,M.
\label{DT2}
\end{equation}

Equations (\ref{DT1}) and (\ref{DT2})  give $M+1$ conditions on the $2M+1$ parameters corresponding to the argument of $\mathcal{C}$ and the points $\{b_{1}^{(j)}, b_{2}^{(j)}|j=1,...,M\}$. 
The other set of $M$  conditions necessary to determine these parameters comes from the requirement that
 $W(\zeta)$ be single-valued in $D_\zeta$. This implies that a $2\pi$-traversal around $C_j$ should correspond to returning to the same point on the $j$-th vertical slit, i.e.,
\begin{equation}
\mathrm{Im}\left[\oint_{C_j}W_{\zeta}(\zeta') d\zeta'\right]=0, \qquad j=1,...,M.
\label{bcs2}
\end{equation}
Thus, once the conformal moduli $q_j$ and $\delta_j$ are specified, the conditions (\ref{DT1}), (\ref{DT2}), and (\ref{bcs2}) give $2M+1$ real equations which can be solved numerically to determine  $\arg(\mathcal{C})$ and the points $\{b_{1}^{(j)}, b_{2}^{(j)}|j=1,...,M\}$. From a numerical viewpoint, however, working with  equations (\ref{DT1}) and (\ref{DT2}) is more cumbersome because one does not know {\it a priori} the values of the   integers $n$ and $m$ (which have to be found by trial and error). 
To circumvent this problem, we enforce the boundary conditions (\ref{DT1a}) and (\ref{DT2a}) on a particular point, say, $\theta=\pi/2$, and then solve these equations, together with (\ref{bcs2}), via a multivariate Newton's method for root finding. (In the examples shown in the next section, the corresponding accessory parameters were obtained using this method.)

The last parameter that  needs to be determined is the modulus of the pre-multiplier $\mathcal{C}$.  This  is obtained from the requirement that the logarithmic singularities of $W(\zeta)$   at $\zeta=\pm1$ have the appropriate strength, so that the jump in $W(\zeta)$ when crossing either one of these singularities equals $2\mathrm{i}$, which corresponds to the channel width in the $w$-plane; see Fig.~\ref{schems} (left panel).  The modulus  $|\mathcal{C}|$ is then found to be
\begin{equation}
|\mathcal{C}|=\frac{2}{\pi}\left|\prod_{j=1}^{M} \frac{\omega(1,\gamma_{1}^{(j)})\omega(1,\gamma_{2}^{(j)})}{\omega(1,b_{1}^{(j)})\omega(1,b_{2}^{(j)})}\right|,
\end{equation}
where we have used that $\omega(\zeta,\gamma)=\zeta-\gamma$, as $\zeta\to\gamma$.
This  completes the construction of the function $W_\zeta(\zeta)$.

\subsection{Conformal map $z(\zeta)$}

In light of definition (\ref{eq:z1}), the conformal map $z(\zeta)$ that we seek will  have the following integral form
\begin{equation}
z(\zeta)=\mathcal{A}+\frac{1}{U} \int_{\zeta_0}^{\zeta} \left[W_\zeta(\zeta')-T_\zeta(\zeta') \right ] d\zeta',
\label{eq:z}
\end{equation}
where $\mathcal{A} \in \Bbb{C}$ is a constant, $\zeta_0 \in \Bbb{C}$ is an arbitrary point inside $D_\zeta$, and expressions for $T_{\zeta}(\zeta)$ and $W_{\zeta}(\zeta)$ are given in (\ref{Wzeta}) and (\ref{Tzeta}), respectively.
Without loss of generality we can  set the bubble velocity to $U=2$; it is demonstrated by Vasconcelos \cite{Giovani1} that all other bubble assemblies corresponding to different values of $U$ can be obtained from the $U=2$ solutions by a simple re-scaling. 

Let us  recall that   we have  used up  two of the three degrees of freedom associated with the Riemann-Koebe mapping theorem, namely, choosing  $\zeta=\pm 1$ to map to the channel ends. The remaining  degree of freedom can now be used to fix the value of the  constant $\mathcal{A}$. We are then left with $3M$ free real parameters  corresponding to the $3M$ conformal moduli of our circular domain $D_\zeta$. Physically, these parameters correspond to the area and centroids for each of the $M$ bubbles. Once the $3M$ conformal moduli are prescribed, we can  determine all other parameters entering  (\ref{eq:z}), namely,  the set of points $\{\gamma_{1}^{(j)},\gamma_{2}^{(j)}, b_{1}^{(j)},b_{2}^{(j)}\in C_j|~j=1,...M\}$ on the inner circles and the pre-multiplier $\mathcal{C}$, and thus a specific solution corresponding to a particular bubble assembly is obtained. In the next section, we will  illustrate the foregoing theory by considering some specific examples of various bubble configurations. 

\section{Examples}

\begin{figure}[t]
\centerline{\includegraphics[scale=0.4]{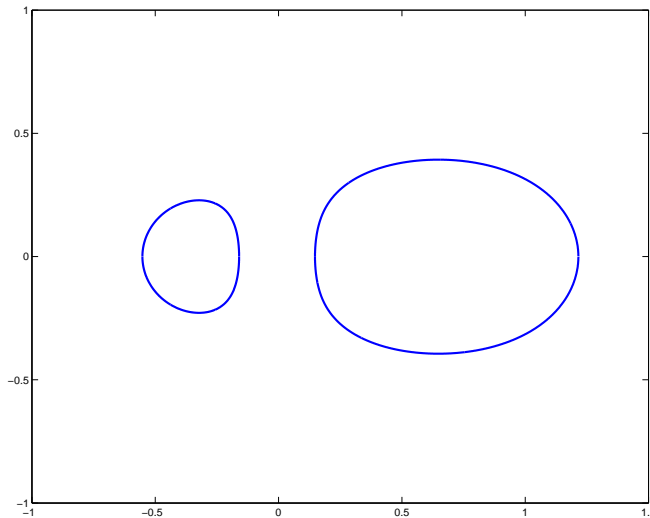}}
\caption{An example of two bubbles which are reflectionally symmetric about the channel centreline. Here the conformal moduli of $D_\zeta$ are $q_1=0.17$, 
$q_2=0.22$, $\delta_1=0$, and $\delta_2=0.6$.}
\label{2bubbles}
\end{figure}

\begin{figure}
\centerline{\includegraphics[scale=0.4]{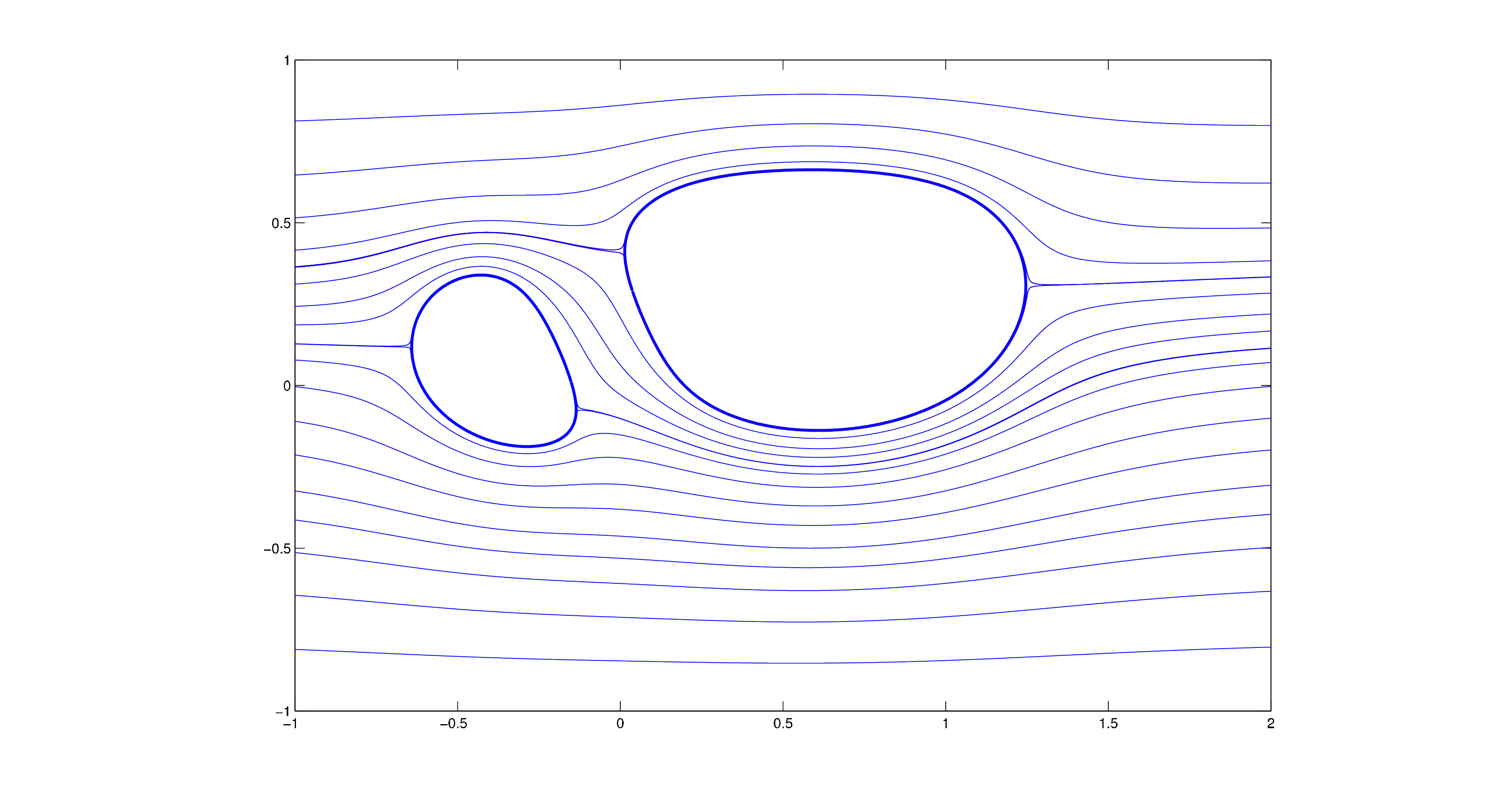}}
\caption{An example of two bubbles in a general asymmetric configuration with a set of  streamlines superposed.}
\label{streamlines}
\end{figure}

Solutions for multiple steady bubbles that either are reflectionally symmetric about the channel centreline or have fore-and-aft symmetry have been obtained by Vasconcelos \cite{Giovani1} by  reducing the problem---on account of the symmetry---to a simply connected domain, and then applying the standard Schwarz-Christoffel formula. These symmetric solutions are readily recovered in our formalism by choosing circular domains $D_{\zeta}$ with the appropriate symmetry: for bubbles with centreline (fore-and-aft) symmetry one must choose a domain $D_{\zeta}$ that is symmetric with respect to reflections about real (imaginary) axis. Figure \ref{2bubbles} shows an example of two bubbles whose centroids are aligned along the channel centreline and which are reflectionally symmetric about it. 
In fact, we were able to successfully recover similar bubbles shapes as in Figure \ref{2bubbles} using the analytical solutions of Vasconcelos \cite{Giovani1}.
 The formalism presented here is however much more general is that it applies to arbitrary bubble configurations, i.e., with no imposed symmetry \emph{a priori}. 
Figure \ref{streamlines} shows an example of two asymmetric bubbles. Several streamlines of flow field in the co-travelling frame  have also been plotted---these streamlines provide a qualitative check on the solutions. 

\begin{figure}
\begin{center}
\centerline{\includegraphics[scale=0.4]{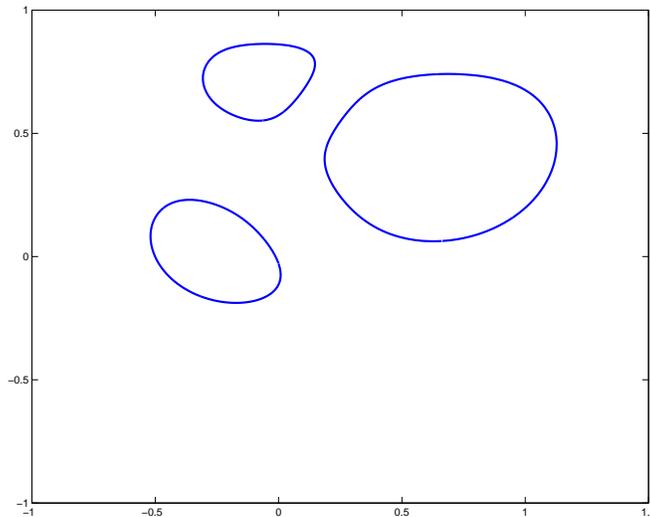}}
\caption{An example of three bubbles in a general asymmetric configuration. For this bubble assembly, the following conformal moduli of $D_\zeta$ were picked: $q_1=0.18, 
q_2=0.22, q_3=0.195, \delta_1=0, \delta_2=0.6+0.23i, \delta_3=0.2+0.63i$.}
\label{3bubbles}
\end{center}
\end{figure}

\begin{figure}[!h]
\begin{center}
\centerline{\includegraphics[scale=0.4]{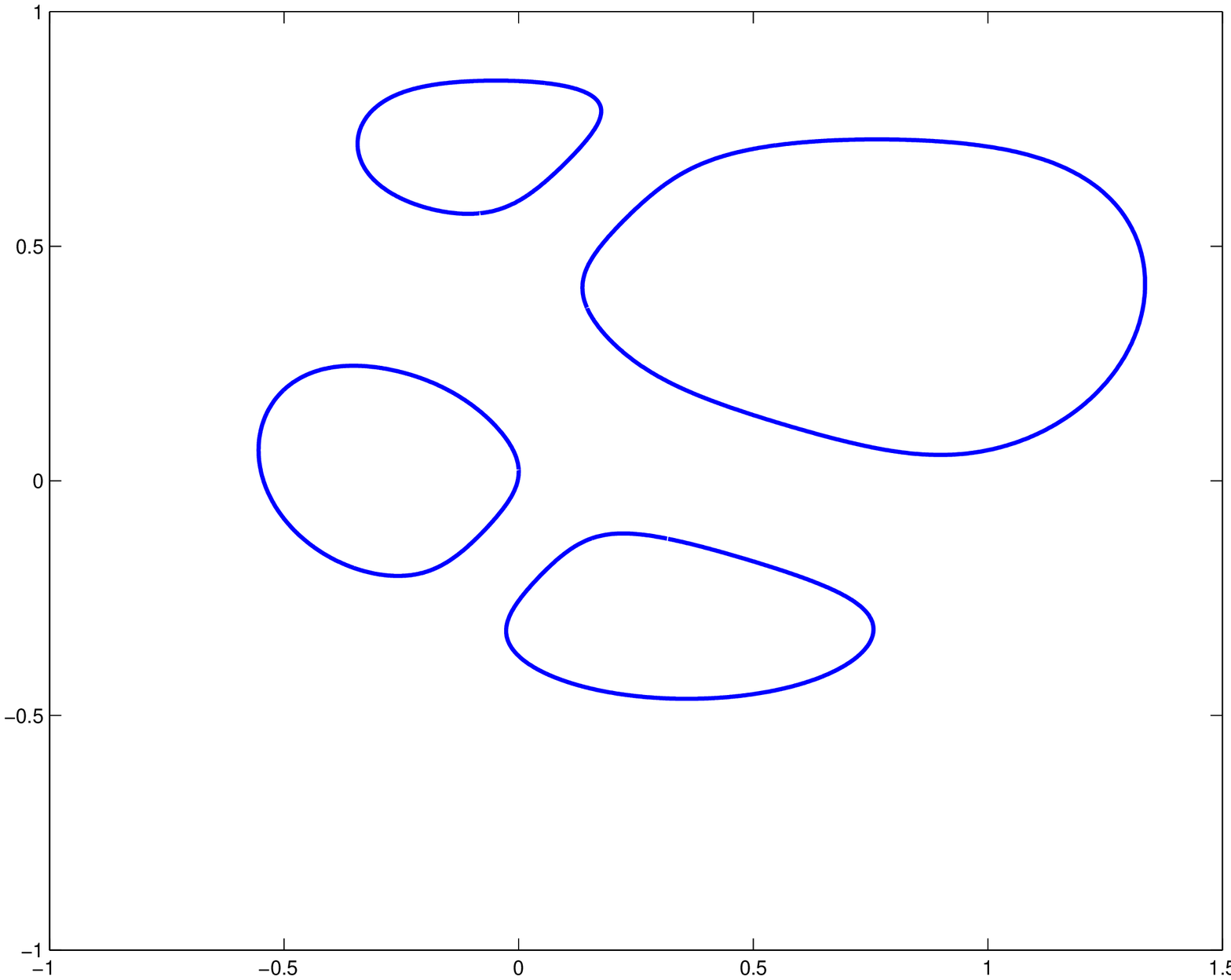}}
\caption{An example of four bubbles in a general asymmetric configuration. We chose the following conformal moduli of $D_\zeta$: $q_1=0.19, q_2=0.235, q_3=0.2, q_4=0.175, \delta_1=0, \delta_2=0.6+0.23i, \delta_3=0.2+0.63i, \delta_4=0.4-0.25i$.}
\label{5bubbles}
\end{center}
\end{figure}

The versatility and generality of our method can be demonstrated through solving for parameters yielding a larger number of bubbles in some asymmetric configuration.  Figure \ref{3bubbles} shows an example of the bubble shapes for a particular asymmetric assembly of three bubbles, whereas Fig.~\ref{5bubbles} reveals the shape of the bubble boundaries in a particular four-bubble configuration. These assemblies are not symmetric about any axis and the bubbles in each assembly all have  different areas (corresponding to the different choices of $q_j$ values).  Solutions for a higher number of bubbles can be treated in a similar manner but the numerical computation of the parameters becomes increasingly more expensive.

In each of the Figs.~\ref{2bubbles}-\ref{5bubbles}
the interaction between the bubbles themselves, and between the bubbles and the channel walls, is clearly visible from the shapes of their boundaries. As specified in the respective figure captions, we have made particular choices of the conformal moduli defining $D_\zeta$ in order to find the finite set of accessory parameters in our conformal map determining the bubble shapes. Alternatively, we could have specified the areas and the centroids of each of the individual bubbles and solved for the conformal moduli of $D_\zeta$. However, this would have been a rather challenging numerical undertaking, owing to the fact that the parameters $\{\gamma_{1}^{(j)},\gamma_{2}^{(j)}\in C_j|~j=1,...,M\}$ (the set of preimages of the end points of the $M$ horizontal slits in the $\tau$-plane) would need to be calculated on each iterative step.

\section{Discussion}

We have presented analytical solutions to the free boundary problem of determining the interface shapes of a finite number $M$ of bubbles steadily translating along a Hele-Shaw channel.
To do this, we found a concise formula in the form of an explicit indefinite integral for the conformal map from a bounded ($M+1$)-connected circular domain to the fluid region in the channel exterior to the $M$ bubbles.
The integrand of this indefinite integral is neatly expressed in terms of products of Schottky-Klein prime functions and their derivatives, and is known up to a finite set of accessory parameters to be found as part of the solution.

Our formula for the conformal map determining the bubble shapes is very general and
makes no \textit{a priori} symmetry assumptions concerning the geometrical arrangement of the bubbles. Indeed, all the geometrical information about the physical domain is encapsulated in the prescription of the preimage domain $D_\zeta$ over which each of the Schottky-Klein prime functions appearing in (\ref{eq:z}) is defined. In previous works on steady bubbles in a Hele-Shaw channel,  symmetry had to be enforced in order to make progress. 
Solutions for a single bubble with centreline symmetry were first found by Taylor \& Saffman \cite{TS}. These solutions were later extended by Tanveer \cite{TanveerBub} to include asymmetric bubbles; an alternative derivation of the Tanveer solutions was subsequently given by Combescot \& Dombre \cite{CD1988}  using Riemann-Hilbert methods.  A general class of exact solutions for multiple steady bubbles with either centreline or fore-and-aft symmetry was obtained by  Vasconcelos \cite{Giovani1} by reducing the problem---on account of symmetry---to a simply connected flow domain. Our solution scheme  readily incorporates the solutions found in these works and so these solutions can be viewed as special cases of ours. It also generalises  the solutions for multiple bubbles in an unbounded Hele-Shaw cell obtained by Crowdy \cite{CrowdyUnboundedHSbubs} by including the effect of the channel walls. Indeed, it is possible to retrieve the solutions of Crowdy \cite{CrowdyUnboundedHSbubs} from those presented here by taking the limit as the channel width becomes infinite. This can be accomplished in our formalism by choosing a circular domain $D_{\zeta}$  where  an inner circle is mapped to the channel walls (with the unit circle mapped  to one of the bubbles) and then taking the radius of this inner circle to zero, but we have not  pursued this detail here.

In conclusion, we have devised a constructive method for finding solutions to the free boundary problem of multiple  bubbles (with no assumed symmetry) in a Hele-Shaw channel. The crucial step in our scheme was to represent the conformal mapping from a circular domain to the physical flow region as a sum of two analytic functions that map the circular domain to  multiply connected degenerate polygonal domains. These analytic functions, which correspond to the complex potentials in the laboratory and moving frames, were then obtained from the generalised Schwarz-Christoffel formula for multiply connected domains in terms of Schottky-Klein prime function. We considered examples of various bubble assemblies, demonstrating that our solution scheme is capable of modelling any finite number of bubbles in a particular assembly.

An interesting extension of the present work would be to consider a doubly periodic array of asymmetric bubble assemblies. This, in turn, would generalise the solutions of Silva \& Vasconcelos \cite{GiovaniSilva} for a doubly periodic array of symmetric bubbles. It is  worth noting that an exact solution for an asymmetric stream of bubbles (with one bubble per unit cell) in a Hele-Shaw channel was recently found by Silva \& Vasconcelos \cite{GiovaniSilva2} using conformal mapping between doubly connected domains. It is possible to extend these solutions using the ideas presented in this paper to include a periodic assembly of bubbles with multiple asymmetric bubbles per unit cell; work is currently in progress in this direction. Another interesting line of enquiry would be to investigate the effects of surface tension on the bubble boundaries. 
This gives rise to the so-called selection problems: for non-zero surface tension, there is no longer a continuum of bubble velocities for which solutions can exist and so only a discrete set of  velocities are allowed. It is also important to point out that obtaining exact analytical solutions for steady Hele-Shaw systems naturally paves the way to finding time-dependent solutions which are also of great physical and mathematical interest. An example of the extension of the steady theory to the time-dependent case is the  recent study by Mineev-Weinstein \& Vasconcelos \cite{MarkGio}, who have been able to determine an exact solution for the time evolution of a bubble of arbitrary initial shape using elliptic functions. 
Our new solutions---in terms of Schottky-Klein prime functions for domains of arbitrary connectivity---should therefore serve as a starting point for  future endeavours looking into constructing time-dependent solutions for multiple bubbles in a Hele-Shaw channel.

\section*{Acknowledgments}
{The authors wish to thank D. G. Crowdy for helpful discussions.  CCG is  appreciative of the hospitality of the Department of Physics at the Federal University of Pernambuco where part of this work was carried out. GLV would like to thank the hospitality of the Department of Mathematics at Imperial College London where this work was completed.}
{CCG acknowledges financial support from a Doctoral Prize Fellowship of the Engineering and Physical Sciences Research Council (United Kingdom). GLV acknowledges financial support from a scholarship from the Conselho Nacional de Desenvolvimento Cientifico e Tecnologico (Brazil) for a sabbatical stay at Imperial College London.}



\begin{thebibliography}{}

\bibitem{Pelce} Pelc\'e P. 1988 {\it Dynamics of Curved Fronts}. Academic Press,
San Diego.

\bibitem{Gustafsson}
Gustafsson B, Vasil'ev A. 2006 {\it Conformal and Potential Analysis
in Hele-Shaw Cell}. Birkh\"auser, Basel. 

\bibitem{Mark1} Mineev-Weinstein M, Putinar M, Teodorescu R.  2008 Random Matrices in 2D, Laplacian Growth and Operator Theory. {\it J. Phys. A} {\bf 41}, 263001. (doi:10.1088/1751-8113/41/26/263001)

\bibitem{Gakhov} Gakhov FD. 1990 {\it Boundary value problems}. Dover, New York.

\bibitem{DarrenRHpaper} Crowdy DG. 2009 Explicit solution of a class of {R}iemann-{H}ilbert problems. {\it Ann. Univ. Paedagog. Crac. Stud. Math.} {\bf 8}, 5-18.

\bibitem{BoundedMCSC} Crowdy DG. 2005 The {S}chwarz-{C}hristoffel mapping to bounded multiply connected polygonal domains. {\it Proc. Roy. Soc. A} {\bf 461}, 2653-2678. (doi: 10.1098/rspa.2005.1480)

\bibitem{UnboundedMCSC} Crowdy DG. 2007 Schwarz-{C}hristoffel mappings to unbounded multiply connected polygonal regions. {\it Math. Proc. Cambridge Philos. Soc.} {\bf 142}, 319-339. (doi: 10.1017/S0305004106009832)

\bibitem{TS} Taylor GI, Saffman PG. 1959 A note on the motion of bubbles in a {H}ele-{S}haw cell and porous medium. {\it Quart. J. Mech. Appl. Math.} {\bf 12}, 265-279. (doi: 10.1093/qjmam/12.3.265)

\bibitem{TanveerBub} Tanveer S. 1987 New solutions for steady bubbles in a {H}ele-{S}haw cell. {\it Phys. Fluids} {\bf 30}, 651. (doi: 10.1063/1.866369)

\bibitem{Giovani1} Vasconcelos GL. 2001 Exact solutions for steady bubbles in a {H}ele-{S}haw cell with rectangular geometry. {\it J. Fluid Mech.} {\bf 444}, 175-198. (doi: 10.1017/S0022112001005365)

\bibitem{GiovaniSilva} Silva AMP, Vasconcelos GL. 2011 Doubly periodic array of bubbles in a {H}ele-{S}haw cell. {\it Proc. Roy. Soc. A} {\bf 467}, 346-360. (doi: 10.1098/rspa.2010.0227)

\bibitem{GiovaniSilva2} Silva AMP, Vasconcelos GL. 2013 Stream of asymmetric bubbles in a {H}ele-{S}haw channel. {\it Phys. Rev. E} {\bf 87}, 055001. (doi: 10.1103/PhysRevE.87.055001)

\bibitem{Giovani2} Vasconcelos GL. 1993 Exact solutions for a stream of bubbles in a {H}ele-{S}haw cell. {\it Proc. Roy. Soc. A} {\bf 442}, 463-468. (doi: 10.1098/rspa.1993.0114)

\bibitem{Giovani3} Vasconcelos GL. 1994 Multiple bubbles in a Hele-Shaw cell. {\it Phys Rev. E} {\bf 50}, R3306-R3309. (doi: 10.1103/PhysRevE.50.R3306)

\bibitem{CrowdyUnboundedHSbubs} Crowdy DG. 2009 Multiple steady bubbles in a {H}ele-{S}haw cell. {\it Proc. Roy. Soc. A} {\bf 465}, 421-435. (doi: 10.1098/rspa.2008.0252)

\bibitem{wrongPaper} Crowdy DG. 2009 An assembly of steadily translating bubbles in a {H}ele-{S}haw channel. {\it Nonlinearity} {\bf 22}, 51-65. (doi: 10.1088/0951-7715/22/1/004)

\bibitem{Goluzin} Goluzin GM. 1969 {\it Geometric theory of functions of a complex variable}. American Mathematical Society, Providence.

\bibitem{Fay} Fay JD. 2008. {\it Theta functions on Riemann surfaces}. Springer, New York.

\bibitem{cmft_compSK}
Crowdy DG, Marshall JS. 2007 Computing the Schottky-Klein prime function on the Schottky double of planar domains. {\em Comput. Methods Funct. Theory} \textbf{1},  293-308. (doi: 10.1007/BF03321646)

\bibitem{SC2013} Vasconcelos GL, Green CC. 2013 A note on the Schwarz-Christoffel mapping for multiply connected polygonal domains (in preparation). 

\bibitem{CD1988} Combescot R,  Dombre T. 1988 Selection in the Saffman-Taylor bubble and asymmetrical finger problem. {\it Phys. Rev. A} {\bf 38}, 2573-2581. (doi: 10.1103/PhysRevA.38.2573)

\bibitem{MarkGio} Mineev-Weinstein M, Vasconcelos GL. 2013 Selection of the {T}aylor-{S}affman bubble does not require surface tension. Submitted to {\it Physical Review Letters}.

\end{thebibliography}
\end{document}